\documentclass[a4paper]{jpconf}

\usepackage{graphicx}
\usepackage{upgreek}
\usepackage{bm}
\usepackage{epstopdf}

\begin{document}

\title{The effect of an offset-dipole magnetic field on the Vela pulsar's $\gamma$-ray light curves}

\author{M Breed$^1$, C Venter$^1$, A K Harding$^2$ and T J Johnson$^3$}

\address{$^1$ Centre for Space Research, North-West University, Potchefstroom Campus, Private Bag X6001, Potchefstroom, 2520, South Africa}
\address{$^2$ Astrophysics Science Division, NASA Goddard Space Flight Center, Greenbelt, MD 20771, USA}
\address{$^3$ National Research Council Research Associate, National Academy of Sciences, Washington, DC 20001, resident at Naval Research Laboratory, Washington, DC 20375, USA}

\ead{20574266@nwu.ac.za}

\begin{abstract}
Over the past six years, the {\it Fermi} Large Area Telescope has detected more than 150 $\gamma$-ray pulsars, discovering a variety of light curve trends and classes. Such diversity hints at distinct underlying magnetospheric and/or emission geometries. We implemented an offset-dipole magnetic field, with an offset characterised by parameters epsilon and magnetic azimuthal angle, in an existing geometric pulsar modelling code which already includes static and retarded vacuum dipole fields. We use these different $B$-field solutions in conjunction with standard emission geometries, namely the two-pole caustic and outer gap models (the latter only for non-offset dipoles), and construct intensity maps and light curves for several pulsar parameters. We compare our model light curves to the Vela data from the second pulsar catalogue of {\it Fermi}. We use a refined chi-square grid search method for finding best-fit light curves for each of the different models. Our best fit is for the retarded vacuum dipole field and the outer gap model.
\end{abstract}

\section{Introduction}

The discovery of the first pulsar in 1967 by Bell and Hewish \cite{Hewish1968} gave birth to pulsar astronomy. Pulsars are believed to be rapidly-rotating, compact neutron stars that possess strong magnetic, electric, and gravitational fields \cite{Abdo2010}. They emit radiation across the entire electromagnetic spectrum, including radio, optical, X-ray, and $\gamma$-rays \cite{Becker2007}. Since the launch of the {\it Fermi Gamma-ray Space Telescope} in June 2008, over 150 $\gamma$-ray pulsars have been detected, of which the Crab and Vela pulsars are the brightest sources. {\it Fermi} consists of two parts including the Large Area Telescope (LAT) and the Gamma-ray Burst Monitor. The LAT measures $\gamma$-rays in the energy range between 20 MeV and 300 GeV \cite{Atwood2009}. Over the past six years, {\it Fermi} has released two pulsar catalogues, both describing the light curve profiles and spectral characteristics of $\gamma$-ray pulsars \cite{Abdo2010,Abdo2013}. The light curves show great variety in profile shape, and may be divided into three general classes based on the relative phase differences between the radio and $\gamma$-ray pulses \cite{Espinoza2013,Venter2009}. The light curves also show energy-dependent behaviour. Most of the young and millisecond pulsars exhibit two $\gamma$-ray peaks whereas some pulsars including Vela display three peaks \cite{Abdo2013}.
\begin{figure}[h!]
	\includegraphics[width=20pc,height=14pc]{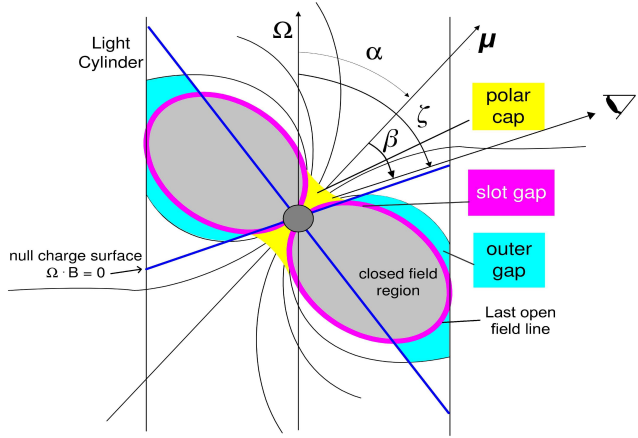}\hspace{0.2cm}
\begin{minipage}[b]{17pc}\caption{\label{models} \small A schematic representation of the different geometric pulsar models. The PC model extends from the neutron star surface up to low-altitudes above the surface (yellow region). The TPC emission region (curved magenta lines) extends from $R_{\rm NS}$ (neutron star radius) up to the light cylinder $R_{\rm LC}$ (where corotation speed equals the speed of light, vertical lines), the OG region (cyan regions) from $R_{\rm NCS}$ (the null charge surface, where the Goldreich-Julian charge density $\rho_{GJ} = 0$ \cite{Goldreich1969}, indicated by the dark blue lines) up to the $R_{\rm LC}$, and the PSPC from $R_{\rm NS}$ to the $R_{\rm LC}$, covering the full open volume region. Adapted from \cite{Harding2004}.}
\end{minipage}
\end{figure}

Models of pulsar geometry are characterised by the inclination angle ($\alpha$) between the rotation ($\boldsymbol{\Omega}$) and magnetic ($\boldsymbol{\mu}$) axes, the observer's viewing angle ($\zeta$) between the rotation axis and the observer's line of sight, and the impact angle ($\beta\equiv\mid\zeta-\alpha\mid$). Geometric models assume the presence of several `gap regions' in the pulsar magnetosphere. These are defined as regions where particle acceleration and emission take place. These geometric models include the two-pole caustic (TPC) \cite{Dyks2003} (the slot gap (SG) \cite{Muslimov2003} model may be its physical representation), outer gap (OG) \cite{Cheng1986,Romani1996}, and pair-starved polar cap (PSPC) model \cite{Harding2005}. All of these models are represented in figure~\ref{models}. 
The emissivity $\epsilon_{\nu}$ of high-energy photons within this gap region is assumed to be uniform in the corotating frame for geometric models (for physical models the $\epsilon_{\nu}$ changes with radial distance when an electric field is assumed). Since the $\gamma$-rays are expected to be emitted tangentially to the local magnetic field in the corotating frame \cite{Dyks2004a}, the assumed magnetic field geometry is very important with respect to the predicted light curves. Several magnetospheric structures have been studied, including the static dipole field \cite{Griffiths1995}, the retarded vacuum dipole field (RVD) \cite{Deutsch1955} and the offset-dipole $B$-field. The latter is motivated by the fact that retardation of the $B$-field at the light cylinder causes offset of the polar caps (PCs), always toward the trailing edge of the PC ($\phi_0 = \pi/2$) \cite{Harding2011}. The offset is characterised by parameters epsilon ($\epsilon$) and magnetic azimuthal angle ($\phi_{0}$) which represents a shift of the PC away from the magnetic axis, with $\epsilon=0$ corresponding to the static-dipole case. 

In this paper we studied the effect of using different combinations of magnetospheric structures, geometric models, and model parameters on $\gamma$-ray light curves. In Section~\ref{section:review} we discuss the implementation of an offset-dipole solution and associated electric field. Section~\ref{section:matching} describes how we matched limiting cases of the low-altitude and high-altitude $E$-fields using a matching parameter $\eta_c$. In Section~\ref{section:fits} we describe the chi-squared ($\chi^2$) method we used to search for best-fit light curves. Our results are given in Section~\ref{section:results} and the conclusions follow in Section~\ref{section:conclusions}.

\section{Implementation of an offset-dipole $B$-field}\label{section:review}

We implemented \cite{SAIP2013} an offset-dipole magnetic field \cite{Harding2011} into an existing geometric pulsar modelling code \cite{Dyks2004a} which already includes static and RVD fields. The implementation involves a transformation of the $B$-field from spherical to Cartesian coordinates, rotating both the $B$-field components and its Cartesian frame through an angle $-\alpha$, thereby transforming the $B$-field from the magnetic frame (${\bf{\hat{z}}^\prime}\parallel{\boldsymbol{\mu}}$) to the rotational frame (${\bf{\hat{z}}}\parallel{\boldsymbol{\Omega}}$).
We extended the range of $\epsilon$ for which we could solve the PC rim (for details, see \cite{Dyks2004a}) by enlarging the colatitude range thought to contain the last open field line (tangent to $R_{\rm LC}$, see figure~\ref{models}). 

\section{Matching parameter}\label{section:matching}
It is important to take the accelerating $E$-field into account (in a physical model) when such expressions are available, since this will modulate the $\epsilon_{\nu}$ in the gap (as opposed to geometric models where we just assume constant $\epsilon_{\nu}$ at all altitudes). We use analytic expressions \cite{Muslimov2004} for a low-altitude and high-altitude SG $E$-field in an offset-dipole magnetic geometry. These are matched at a critical scaled radius $\eta_c=r_c/R_{\rm NS}$ to obtain a general $E$-field valid for all altitudes \cite{Muslimov2004}. In previous work we chose $\eta_c=1.4$. In this paper we solve $\eta_{\rm c}$ on each field line. Using the general $E$-field we could solve the particle transport equation \cite{Sturner1995,Daugherty1996} (taking only curvature losses into account) to obtain the particle Lorentz factor $\gamma$ that is necessary for calculating the curvature emissivity.   

\section{Finding best-fit light curves}\label{section:fits}

We tested several fitting methods in order to select the most suitable one. We decided to use the standard $\chi^2$ method. For each combination of $B$-field and geometric model, we calculate $\chi^2$ for each set of free model parameters $\alpha$, $\zeta$, normalisation ($A$) and phase shift ($\Delta\phi$). We assume that for the bright Vela pulsar, with a large amount of counts in each bin, the $\chi^2$ will be Gaussian distributed with $N_{\rm dof}=96$ the number of degrees of freedom. For the standard Gaussian distribution, assuming very small Gaussian errors, the reduced $\chi^2_{\nu}$ values are very large and therefore we needed to scale the $\chi^2$ values with the optimal $\chi^2_{\rm opt}$ and multiply by $N_{\rm dof}$. The scaled $\chi^2$ is denoted by $\xi^2$, and the reduced scaled $\chi^2$ will have a value of $\xi^2_{\nu}=1$ \cite{Pierbattista2014}. We next determined confidence intervals (1$\sigma$, 2$\sigma$, and 3$\sigma$) in ($\alpha$,$\zeta$) space for these model parameters, using  $N_{\rm dof}=2$. We only fitted the $\gamma$-ray light curve, because we do not want to bias our results with a simplistic radio model.   

\section{Results}\label{section:results}
\subsection{Phaseplots and light curves}
\begin{figure}[h!]
\begin{minipage}{16pc}
\centering
	\includegraphics[width=14pc,height=21pc]{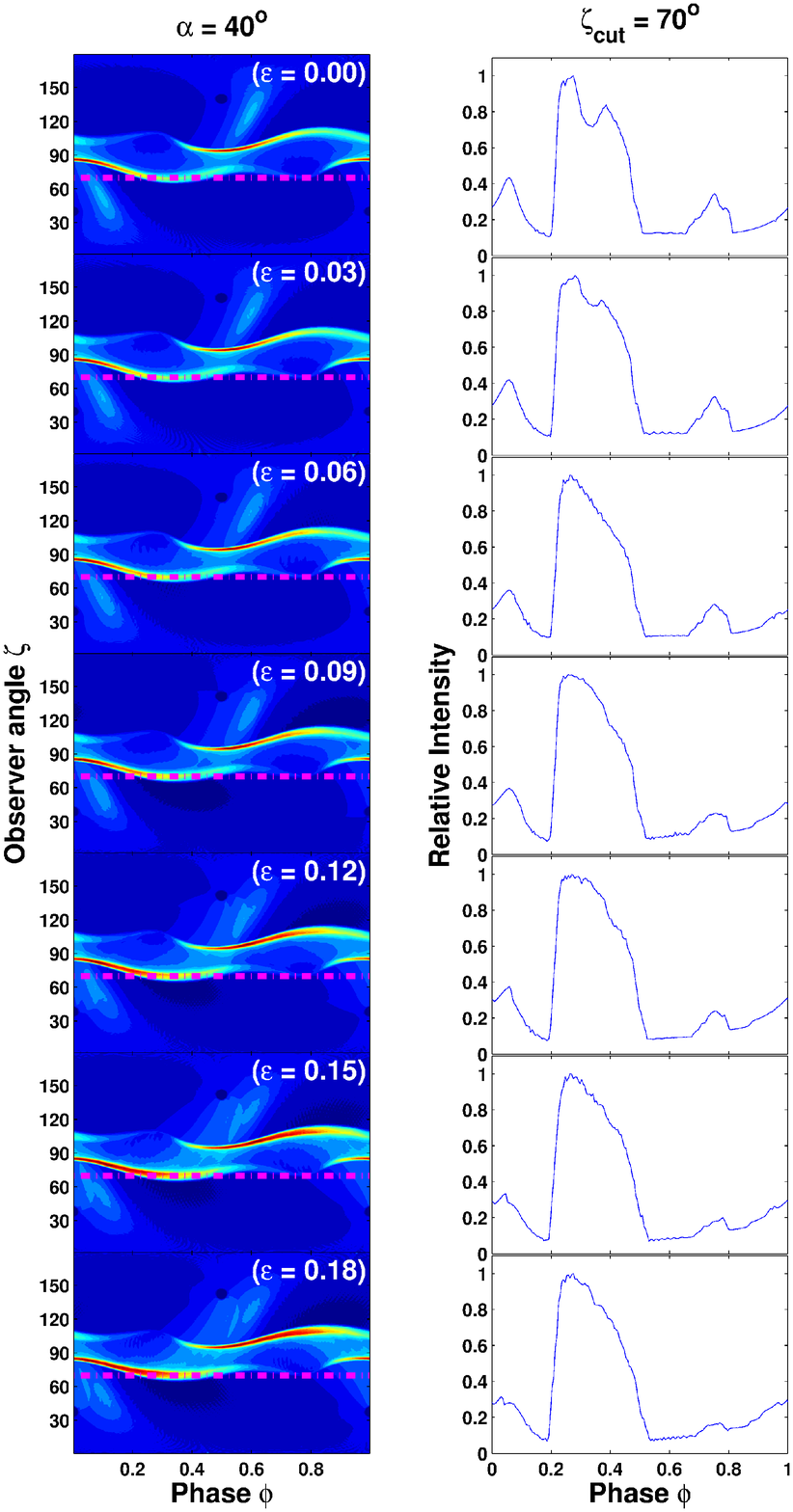}
	\caption{\label{atlas1} \small Phaseplots and light curves for different $\epsilon$ values, for constant $\epsilon_{\nu}$.}
\end{minipage}
\hspace{0.3cm}
\begin{minipage}{16pc}
\centering
	\includegraphics[width=14pc,height=21pc]{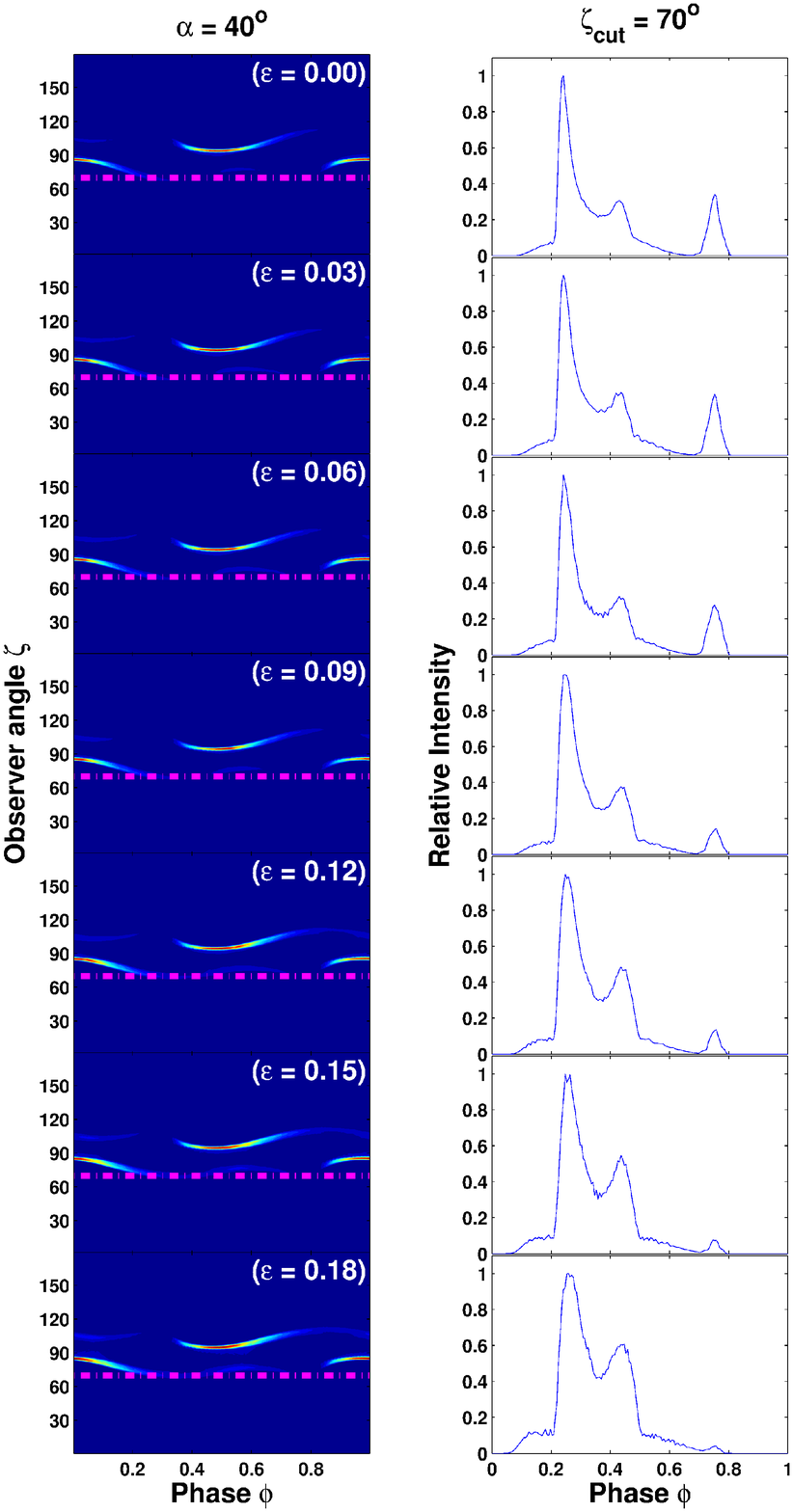}
	\caption{\label{atlas2} \small Phaseplots and light curves for different $\epsilon$ values, for variable $\epsilon_{\nu}$.}
\end{minipage} 
\end{figure}

\begin{figure}[h!]
\centering
	\includegraphics[width=27pc]{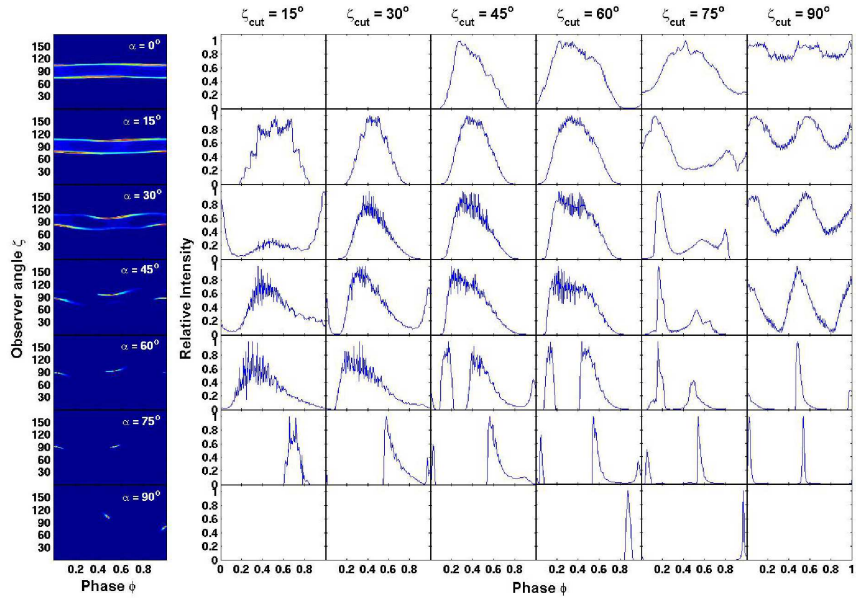}\hspace{0.2cm}
	\caption{\label{atlas3} \small Phaseplots and light curves for different values of $\alpha$ and $\zeta_{\rm cut}$ for a fixed value of $\epsilon=0.18$, for variable $\epsilon_{\nu}$.}
\end{figure}
In figure~\ref{atlas1}, ~\ref{atlas2} and ~\ref{atlas3}, we show the intensity maps or phaseplots (emission per solid angle versus $\zeta$ and pulse phase $\phi$) and their corresponding light curves (i.e., cuts at constant $\zeta=\zeta_{\rm cut}$) for the offset-dipole $B$-field and the TPC model. The dark circle in figure~\ref{atlas1} is the non-emitting PC, and the sharp, bright regions are the emission caustics, where radiation is bunched in phase due to relativistic effects. Figure~\ref{atlas1} and ~\ref{atlas2} represent phaseplots for a fixed $\alpha=40^\circ$ and $\zeta_{\rm cut}=70^\circ$, with $\epsilon$ ranging from 0 to 0.18 with increments of 0.03. We contrasted the cases of constant and variable $\epsilon_{\rm \nu}$. We observed a qualitative difference in caustic structure. The caustics seem larger and more pronounced in the constant $\epsilon_{\nu}$ case. In figure~\ref{atlas3} we chose a fixed value of $\epsilon=0.18$ for variable $\epsilon_{\nu}$, with $\alpha$ ranging from 0$^\circ$ to 90$^\circ$ and $\zeta_{\rm cut}$ from 15$^\circ$ to 90$^\circ$, both with a resolution of 15$^\circ$. This shows examples of various light curves that may be obtained in this model.  

\subsection{Contours and best-fit light curves}

\begin{figure}[h!]
\centering
	\includegraphics[width=32pc,height=31pc]{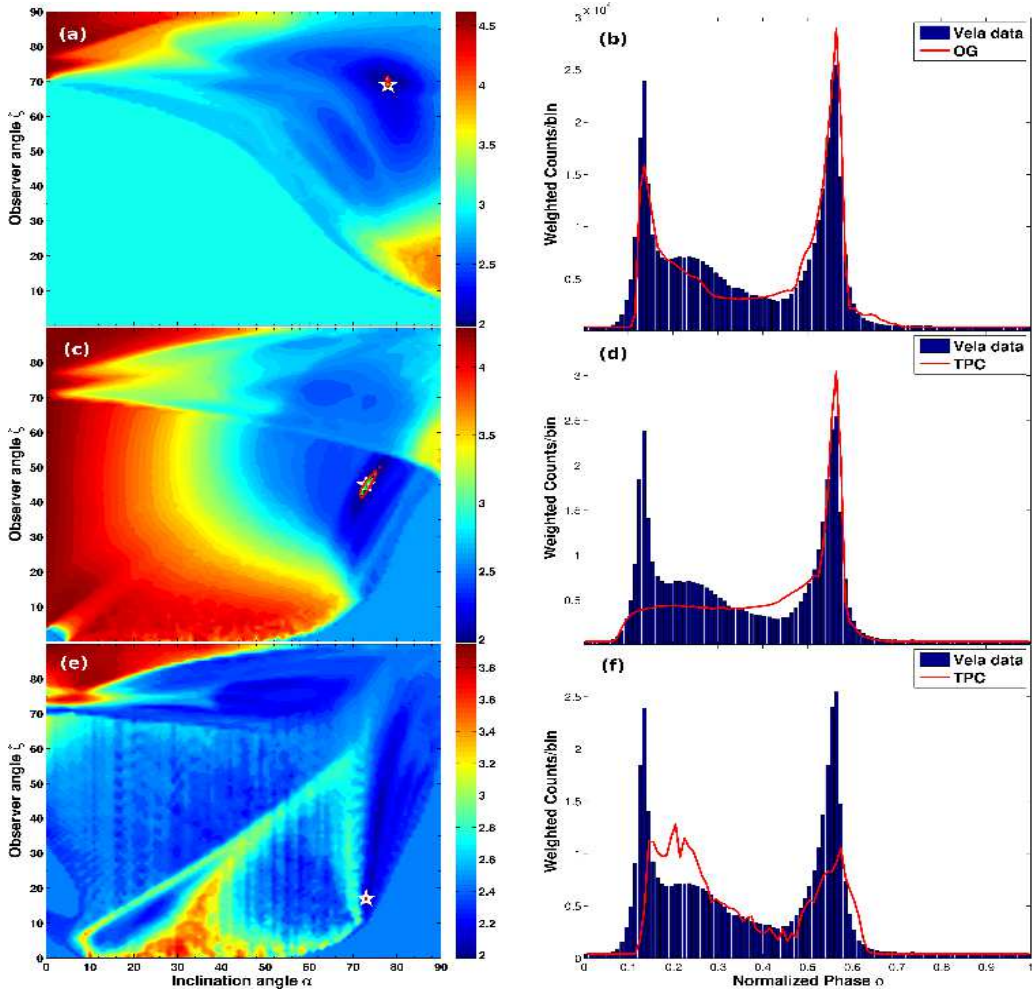}
{\small \caption{\label{bestfits} Contour plots (left) and corresponding best-fit light curves (right). Panels (a) and (b) represent the best fit for the RVD and the OG model. Panels (c) and (d) are for the offset dipole and TPC model for constant $\epsilon_{\nu}$ and $\epsilon=0.00$. Panels (e) and (f) are for the offset dipole and TPC model for variable $\epsilon_{\nu}$ and $\epsilon=0.18$. The colour bar of the contour plots represent log$_{\rm 10}\xi^2$. The confidence contour for 1$\sigma$ (magenta line), 2$\sigma$ (green line), and 3$\sigma$ (red line) are also shown. The star indicates the best-fit solution. The blue histogram denotes the observed Vela profile (for energies $>100$ MeV) \cite{Abdo2013}  and the red line the geometric model.}}
\end{figure}
In figure~\ref{bestfits} we represent our best fits we obtained using the $\chi^2$ method. Our overall best fit is for the RVD field and the OG model, with $\alpha=78^{\circ}$, $\zeta=69^{\circ}$, $A=1.3$, and $\Delta\phi=0$. For the offset-dipole solution and the TPC model, we have a best fit (assuming constant $\epsilon_{\nu}$) for parameters $\epsilon=0.00$, $\alpha=73^{\circ}$, $\zeta=45^{\circ}$, $A=1.3$, and $\Delta\phi=0.55$. The best fit for the offset-dipole solution and the TPC model, assuming variable $\epsilon_{\nu}$, is for parameters $\epsilon=0.18$, $\alpha=73^{\circ}$, $\zeta=17^{\circ}$, $A=0.5$, and $\Delta\phi=0.60$. The best-fit parameters for each $B$-field and geometric model combination are summarised in table~\ref{summary}. The table includes the different model combinations, the optimal $\chi^2$ value (before scaling), the free parameters with errors (found using 3$\sigma$ connected ($\alpha$,$\zeta$) contours), a reference fit found using radio polarisation data, and the comparison between models using the difference between the respective optimal values of $\xi^2$, represented by $\Delta\xi^2_*$.

\begin{table}[h]
{\tiny
{\small \caption{\label{summary} Best-fit parameters for each $B$-field and geometric model combination.}}
\begin{center}
\lineup
\begin{tabular}{*{12}{l}}
\br                              
\multicolumn{2}{l}{Model} & \multicolumn{5}{c}{Our model parameters} & \multicolumn{2}{c}{Ref. fit \cite{Watters2009}} & \multicolumn{2}{c}{Radio pol. \cite{Johnston2005}} & \multicolumn{1}{c}{$\Delta\xi^2_{*}$} \cr 
Combinations &\0\0 $\epsilon$ & $\chi^2$ & $\alpha$ & $\zeta$ & $A$ & $\Delta\phi$ & $\alpha$ & $\zeta$ & $\alpha$ & $\zeta$ & \cr 
    & & ($\times{10}^5$) & ($^\circ$) & ($^\circ$) &  & & ($^\circ$) & ($^\circ$) & ($^\circ$) & ($^\circ$) & \cr 
\mr
\multicolumn{3}{l}{\bf Static dipole:} & & & & & & & & & \cr
TPC & \0\0--- & \0\00.819 & 73$^{+3}_{-2}$ & 45$^{+4}_{-4}$ & 1.3 & 0.55 & & & & & \0\00.000\cr
OG & \0\0--- & \0\00.891 & 64$^{+5}_{-3}$ & 86$^{+1}_{-1}$ & 1.3 & 0.05 & & & & & \0\08.439\cr
\mr
\multicolumn{3}{l}{\bf RVD:} & & & & & & & & & \cr
TPC & \0\0--- & \0\03.278 & 54$^{+5}_{-5}$ &67$^{+3}_{-2}$ & 0.5 & 0.05 & 62--68
& 64 & & & 723.5\cr
OG  & \0\0--- & \0\00.384 & 78$^{+1}_{-1}$ & 69$^{+2}_{-1}$ & 1.3 & 0.00 & 75 & 64 & & & \0\00.000\cr
\mr
\multicolumn{3}{l}{\bf Offset dipole - constant $\epsilon_{\nu}$:} & & & & & & & & & \cr
TPC & \0\0$0.00$ & \0\00.819 & 73$^{+3}_{-2}$  & 45$^{+4}_{-4}$  & 1.3 & 0.55 & & & & & \0\00.000\cr
    & \0\0$0.03$ & \0\00.834 & 73$^{+2}_{-2}$ & 43$^{+4}_{-5}$ & 1.3 & 0.55 & & & & & \0\01.758\cr
    & \0\0$0.06$ & \0\00.867 & 73$^{+2}_{-2}$ & 42$^{+5}_{-5}$ & 1.3 & 0.55 & & & & & \0\05.626\cr
    & \0\0$0.09$ & \0\00.882 & 73$^{+1}_{-2}$ & 41$^{+3}_{-5}$ & 1.3 & 0.55 & & & & & \0\07.385\cr
    & \0\0$0.12$ & \0\01.00 & 74$^{+1}_{-3}$ & 42$^{+3}_{-6}$ & 1.4 & 0.55 & & & & & \021.216\cr
    & \0\0$0.15$ & \0\00.948 & 73$^{+1}_{-2}$ & 39$^{+3}_{-5}$ & 1.4 & 0.55 & & & & & \015.121\cr
    & \0\0$0.18$ & \0\00.969 & 73$^{+2}_{-3}$ & 37$^{+4}_{-4}$ & 1.3 & 0.55 & & & & & \017.582\cr
\mr
\multicolumn{3}{l}{\bf Offset dipole - variable $\epsilon_{\nu}$:}  & & & & & & & & & \cr
TPC & \0\0$0.00$ & \0\01.46 & 21$^{+2}_{-3}$ & 71$^{+1}_{-1}$ & 0.5 & 0.85 & & & & & \017.032\cr
    & \0\0$0.03$ & \0\01.63 & 22$^{+3}_{-2}$ & 71$^{+1}_{-1}$ & 0.5 & 0.85 & & & & & \030.194\cr
    & \0\0$0.06$ & \0\01.68 & 73$^{+1}_{-1}$ & 16$^{+1}_{-3}$ & 0.6 & 0.55 & & & & & \034.065\cr
    & \0\0$0.09$ & \0\01.65 & 73$^{+1}_{-1}$ & 15$^{+1}_{-1}$ & 0.6 & 0.55 & & & & & \031.742\cr
    & \0\0$0.12$ & \0\01.59 & 73$^{+1}_{-1}$ & 14$^{+2}_{-1}$ & 0.7 & 0.55 & & & & & \027.097\cr
    & \0\0$0.15$ & \0\01.52 & 73$^{+1}_{-1}$  & 15$^{+3}_{-2}$  & 0.5 & 0.60 & & & & & \021.677\cr
    & \0\0$0.18$ & \0\01.24 & 73$^{+1}_{-1}$  & 17 $^{+1}_{-1}$ & 0.5 & 0.60 & & & & & \0\00.000\cr
\mr
\multicolumn{2}{l}{} & & & & & & & & 53 & 59.5 & \cr
\br
\end{tabular}
\end{center}
}
\end{table}

\section{Conclusions}\label{section:conclusions}

We have studied the effect of implementing the offset-dipole $B$-field on $\gamma$-ray light curves for the TPC geometry. We find an optimal best-fit for Vela for the RVD $B$-field and the OG model. The OG model displays reduced off-peak emission. We note that the best fits for the offset-dipole $B$-field for constant $\epsilon_{\nu}$ favour smaller values of $\epsilon$ and for variable $\epsilon_{\nu}$ larger $\epsilon$ values. When including an $E$-field the resulting phaseplots becomes qualitatively different compared to constant $\epsilon_{\nu}$. In future, we want to continue to produce light curves using improved geometric models and $B$-fields, and also using more data, in order to search for best-fit profiles, thereby constraining the low-altitude magnetic structure and system geometry of several bright pulsars.

\ack This work is supported by the South African National Research Foundation. AKH acknowledges the support from the NASA Astrophysics Theory Program. CV, TJJ, and AKH acknowledge support from the {\it Fermi} Guest Investigator Program.

\end{document}